\documentclass[superscriptaddress,twocolumn,showpacs,preprintnumbers,amsmath,amssymb]{revtex4}

\usepackage{bm,dcolumn,amsmath,graphicx}

\newcommand{\be}{\begin{equation}}
\newcommand{\ee}{\end{equation}}
\newcommand{\ket}[1]{\left| #1 \right\rangle}
\newcommand{\sub}[1]{_{\mbox{\scriptsize #1}}}
\renewcommand{\vec}[1]{{\bf #1}}

\begin{document}

\title{High precision measurement of the $^{87}$Rb D-line tune-out wavelength}

\author{R. H. Leonard}
\affiliation{Physics Department, University of Virginia, Charlottesville, Virginia 22904, USA}
\author{A. J. Fallon}
\affiliation{Physics Department, University of Virginia, Charlottesville, Virginia 22904, USA}
\author{C. A. Sackett}
\affiliation{Physics Department, University of Virginia, Charlottesville, Virginia 22904, USA}
\author{M. S. Safronova}
\affiliation{Department of Physics and Astronomy, University of Delaware, Newark, Delaware 19716, USA}
\affiliation{Joint Quantum Institute, National Institute of Standards and Technology and the University of Maryland, College Park, Maryland, 19716 USA}

\date{\today}

\begin{abstract}
We report an experimental measurement of a light wavelength
at which the ac electric polarizability equals zero for $^{87}$Rb atoms 
in the $F=2$ ground hyperfine state. 
The experiment uses a condensate
interferometer both to find this `tune-out' wavelength and to accurately
determine the light polarization for it.
The wavelength lies between the D1 and D2 spectral
lines at 790.032388(32) nm.  The measurement
is sensitive to the tensor contribution to the polarizability,
which has been removed so that the reported value is the zero of the
scalar polarizability. The precision is fifty 
times better than previous tune-out wavelength measurements.
Our result can be used to determine
the ratio of matrix elements $\left|
\langle 5P_{3/2}||d||5S_{1/2}\rangle
/\langle 5P_{1/2}||d||5S_{1/2}\rangle\right|^2 = 1.99221(3)$, a 100-fold 
improvement over previous experimental
values. New theoretical calculations for 
the tune-out wavelength and matrix
element ratio are presented. The results are consistent with the experiment,
with uncertainty estimates for 
the theory about an order of magnitude larger than
the experimental precision.
\end{abstract}


\maketitle

\section{Introduction}

The energy shift experienced by an atom in an off-resonant optical field 
has found numerous applications in atom trapping,
manipulation, and measurement. The light shift can be characterized
by a frequency-dependent polarizability, which itself depends 
in detail on the wave function of the electrons in the atom. 
Accurate measurements of the polarizability can therefore be used
to test atomic theory calculations, or as phenomenological inputs to
improve those calculations. Polarizability measurements have a long
history of improving our knowledge of atoms in this way
\cite{Gould05,Mitroy10}.

Precise measurements of the polarizability at optical frequencies are
technically difficult, because the light shift depends also on the
optical intensity and it is hard to accurately determine
the intensity {\em in situ}. However, it is possible to instead
measure a light wavelength at which the polarizability equals zero
\cite{LeBlanc07,Arora11,Holmgren12}.
Since these tune-out wavelengths are independent of the 
intensity, they can be 
accurately measured by 
various methods \cite{Lamporesi10,Holmgren12,Herold12,Henson15,Clark15}.

Tune-out wavelengths can be useful for 
applications involving species-specific optical manipulation
\cite{LeBlanc07,Lamporesi10,Arora11,Schneeweiss14} and optical
Feshbach resonances \cite{Clark15}. 
In addition, it was recently shown 
that tune-out wavelengths can be used with an 
atom interferometer for sensitive detection of rotations and accelerations
\cite{Trubko15}. Improved knowledge of tune-out wavelengths
can lead to better performance in all these applications.

In this paper we report measurement of the tune-out wavelength for 
$^{87}$Rb near 790 nm, with an accuracy of about 30 fm.
This can be compared to the 1.5 to 2 pm precision of 
previously reported values
for this \cite{Lamporesi10} or other tune-out wavelengths 
\cite{Holmgren12,Herold12,Henson15}.
Our result determines 
the ratio of the D-line dipole matrix elements to an accuracy of 15 ppm,
about a factor of 100 better than previously known
\cite{Volz96,Simsarian98,Gutterres02}.
At our precision the measurement is sensitive to many 
new effects including hyperfine interactions \cite{Kien13},
QED effects \cite{Flambaum05}, the Breit interaction \cite{Dzuba06}, 
and the details of the
atomic core and core-valence interactions \cite{Arora11}. The theoretical tools
required to handle these challenges are closely related to those needed for
interpreting results such as atomic parity violation and electric dipole
measurements in terms of fundamental particle properties \cite{Dzuba12}. 
Related calculations are also useful for constraining black-body radiation
shifts in atomic clocks \cite{Safronova12}.
Our measurement can thus serve as a useful test for theories, or could be
taken as a phenomenological input value for improved results.

\section{Experimental Method}

For an alkali atom in state $i$, the polarizability can be expressed 
as
\be
\alpha_i(\omega) = \frac{1}{\hbar}
\sum_f \frac{2\omega_{if}}{\omega_{if}^2-\omega^2} 
\left|d_{if}\right|^2 + \alpha_c + \alpha_{cv}
\label{alphaparts}
\ee
where the sum is over all excited states $f$ of the valence electron.
The transition frequency between $i$ and $f$ is $\omega_{if}$ 
and $d_{if} = \langle f | \vec{d}\cdot\hat{\epsilon}|i\rangle$ 
is the dipole matrix element between $i$ and $f$ for light with polarization
vector $\hat{\epsilon}$. The $\alpha_c$ term is the 
polarizability contribution
from the core electrons while $\alpha_{cv}$ expresses the effect of
core-valence interactions \cite{Arora11}. 
At most frequencies, $\alpha_c$ and $\alpha_{cv}$ are 
small compared to the valence contribution.
However, tune-out wavelengths occur between pairs of states where 
the valence contributions largely cancel. Figure 1(a) shows the
tune-out wavelength between the D1 and D2 lines of Rb.

\begin{figure}
\includegraphics[width=3.5in]{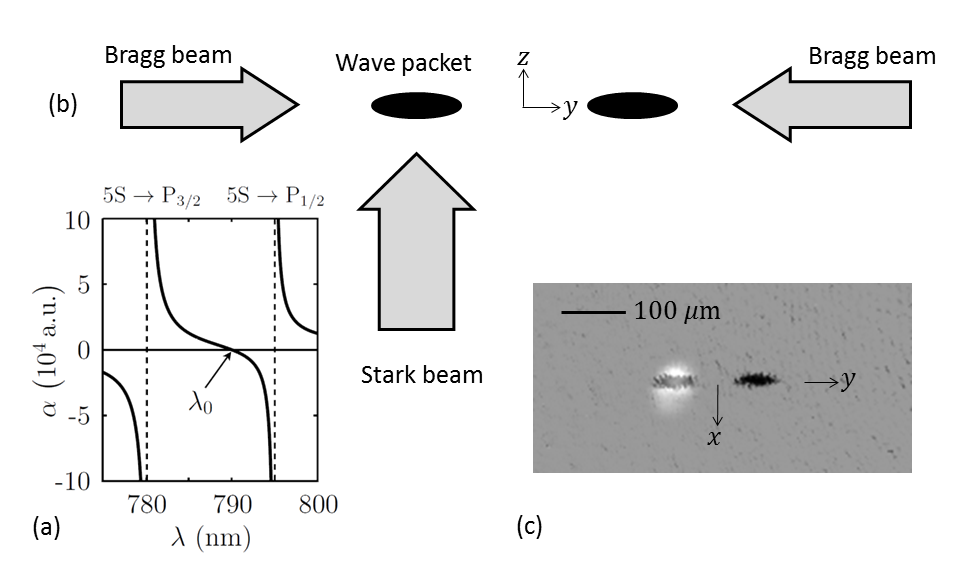}
\caption{
Schematic of measurement. (a) Theoretical plot of the polarizability $\alpha$
for $^{87}$Rb near the D1 and D2 transitions. The polarizability crosses
zero at the tune-out wavelength $\lambda_0$. (b) Optical schematic
for the experiment. The two Bragg laser beams form a standing wave
that is used to split and recombine a Bose condensate to form an
atom interferometer. The Stark laser beam illuminates one of the wave packets
in the interferometer to produce a phase shift. (c) Composite image
of the atomic wave packets (dark) and the Stark beam (white).
Here the wave packet centers are 130~$\mu$m apart after 5 ms of propagation. 
In (b) and (c), 
the coordinate axes $x$, $y$ and $z$ are illustrated.}
\end{figure}

Our measurement uses a Bose condensate atom interferometer, similar to that
previously described in \cite{Burke08}. 
A condensate of about $10^4$ $^{87}$Rb atoms
is produced and loaded into a weak magnetic trap with harmonic
oscillation frequencies of 5.1, 1.1, and 3.2 Hz
along the $x$, $y$, and $z$
directions, respectively. The trap uses a time-orbiting potential,
with a bias field of 20.0 G rotating in the $xz$ plane at 12 kHz
frequency. Oscillating magnetic gradients provide support against
gravity as well as trap confinement.

The atom interferometer is implemented
using an off-resonant standing-wave laser propagating
along the $y$ axis, having wave number $k$. Via Bragg scattering, 
a short pulse from this beam can split the atoms into
two wave packets traveling with momentum $\pm 2\hbar k$
\cite{Hughes07}. After 10 ms, the wave
packets are reflected using another pulse of the Bragg laser, now
adjusted to drive the $\ket{+2\hbar k} \leftrightarrow \ket{-2\hbar k}$
transition. After 20 ms a second reflection pulse is applied,
and after another 10 ms, a recombination pulse is applied. 
By using this
symmetric trajectory, both packets traverse identical paths
in the trap, which reduces phases shifts and fidelity loss from the
trapping potential \cite{Burke08}. 

The recombination pulse brings a fraction $N_0/N$ of the atoms back to rest
in the center of the trap. We obtain $N_0/N = [1+V\cos(\phi+\phi_r)]/2$,
where $\phi$ is the phase difference developed by the atoms during their
separation, $\phi_r$ is the phase shift of the recombination pulse
relative to the initial splitting pulse, and $V \approx 0.7$  
is the visibility.
We here set $\phi_r = \pi/2$ to maximize the sensitivity to $\phi$. 
We measure $N_0/N$ by allowing the three output wave packets to separate for 
40 ms and then observing them via absorption imaging. 

To obtain the polarizability $\alpha$, we focus another laser beam,
traveling along $z$, 
onto one arm of the interferometer. This
Stark beam is applied for 20 ms at the start of the interferometer, 
so that one packet passes through it twice.
Figure 1(b) shows the orientation of the beams involved, and Fig.~1(c)
is a composite image of the atoms and Stark beam together.

\begin{figure}
\includegraphics[width=3.5in]{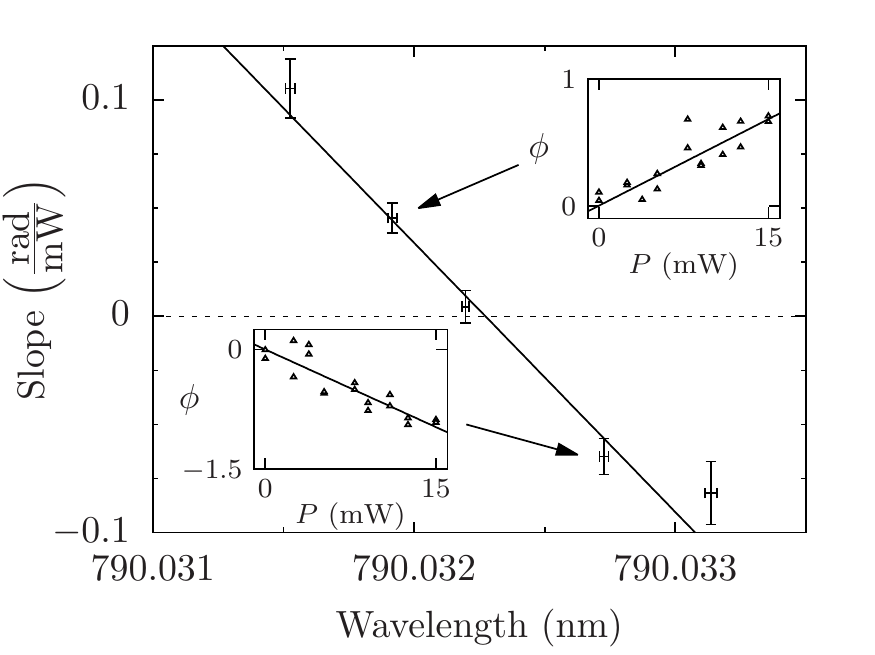}
\caption{
Sample data. The two inset graphs show interferometric measurements of the 
phase shift $\phi$ induced by the Stark beam with power $P$. 
The triangles show individual measurements, which are fit to a line to
determine the slope.
The large graph shows how the slope varies
as a function of the Stark laser wavelength, with the inset graphs
corresponding to the indicated points.
The vertical
error bars are the linear regression errors from the slope fits. The
horizontal error bars are the standard deviation of several wavelength
measurements made over the course of the slope measurement. The line in the
large graph is another linear fit, and the intercept is taken as our
measurement result for the tune-out wavelength $\lambda_0$. Here 
$\lambda_0 = 790.03232$ nm, with a regression error of 50 fm. 
}
\end{figure}

The energy shift $U$ due to the Stark beam is 
\be
U=-\frac{1}{2} \alpha\left\langle \mathcal{E}^2 \right\rangle = 
-\frac{\alpha I}{\epsilon_0 c}
\ee
where $\mathcal{E}$ is the electric field of the beam, 
$I$ is the intensity,  and
$c$ is the speed of light.
The brackets denote time averaging of the optical field. 
The light shift induces a phase 
$\phi = -(1/\hbar) \int U dt$
proportional to the integrated intensity experienced by the atoms. 
We use an approximately Gaussian beam with waist $w \approx 30~\mu$m. 
For a Stark beam power of $P$, this yields
$\phi/(\alpha P) \approx 66$~rad/W for $\alpha$ in atomic units.

The Stark power can be varied from zero to 15 mW using an acousto-optic
modulator. The basic experimental procedure is to 
set the Stark laser to a given wavelength $\lambda$ and run the interferometer
for different beam powers. The resulting phase is fit to a line to determine
the slope, as shown in the Fig. 2 insets. By performing the experiment
at different wavelengths, we plot the slope as a function of $\lambda$.
A second linear fit yields the wavelength $\lambda_0$ at which
the slope and thus $\alpha$ equals zero.

\section{Light Polarization Effects}

A major complication is that $\alpha$ depends strongly
on the optical
polarization of the Stark beam and the orientation of the atomic spin.
In general the energy shift can be expressed as \cite{Kien13}
\be
\label{alpha_terms}
\begin{split}
U = -&\frac{ \langle E^2 \rangle}{2}
\bigg\{ 
\alpha^{(0)} -\mathcal{V}\cos\chi \frac{m_F}{2F} \alpha^{(1)}   \\
&+\left[\frac{3\cos^2\xi -1}{2}\right]
\frac{3m_F^2-F(F+1)}{F(2F-1)}\alpha^{(2)}
\bigg\} \\
\end{split}
\ee
where the $\alpha^{(i)}$ are irreducible components of the
polarizability, namely the scalar $(i=0)$, vector $(i=1)$ and tensor
$(i=2)$ parts. The atom is assumed to be in a particular hyperfine state
$\ket{F,m_F}$ relative to the trap magnetic field direction $\hat{b}
= \vec{B}/B$.
Here we have
$F=m_F=2$. The angle between the Stark beam wave vector
and the magnetic field is $\chi$, so $\cos\chi=\hat{k}\cdot\hat{b}$.
Similarly, $\cos\xi = \hat{\epsilon}\cdot\hat{b}$ is the 
projection of the light polarization vector $\hat{\epsilon}$
onto the magnetic field.
Finally $\mathcal{V}$ is the fourth Stokes parameter for the light, 
characterizing the degree of circular polarization and expressible as
$\mathcal{V}\cos\chi = i(\hat{\epsilon}^*\times\hat{\epsilon})\cdot
\hat{b}$.

We are primarily interested in the scalar polarizability $\alpha^{(0)}$.
The tensor contribution is small but measurable, and will be discussed below.
However, the vector contribution can be quite large.
For instance, for $\sigma_+$ polarized light ($\mathcal{V} = -1$
and $\chi=0$) the vector term
completely eliminates the tune-out wavelength between the D1 and D2 transitions,
since the light does not couple our ground state to any states
in the D1 manifold. To measure the tune-out wavelength of the scalar
term with the desired accuracy, it is necessary to keep $|\mathcal{V}\cos\chi|
< 10^{-5}$. This is challenging since it is comparable to the performance
of the best linear polarizers, and much below the level of polarization
that can typically be maintained when a laser beam passes through
a vacuum chamber window. 

We use two methods to control the vector shift. First, the rotating
bias field of the TOP trap causes $\cos\chi$ in Eq.~\eqref{alpha_terms} to
alternate sign, with a time average close to zero. We verified that
the measurement results did not depend on the phase of the TOP field
at the start of the interferometer.

Second, we linearized the light polarization 
using the interferometer itself. 
Prior to taking a data set such as in Fig.\ 2, we ran the experiment
with the Stark beam pulsed on and off synchronously with the TOP field.
In this way the $\cos\chi$ term could be made close to +1 or -1. We 
adjusted the light polarization so that the measured phase shifts for
those two cases were equal.
The polarization was established with a calcite polarizer, a zero-order
half wave plate, and a zero-order quarter wave plate.
The wave plates could be set to an accuracy of about 0.1$^\circ$,
corresponding to $\mathcal{V} \approx 2\times 10^{-3}$.

After taking the data set, 
the polarization check was repeated and any
difference between the $\cos\chi=\pm 1$ phases was used to estimate the 
polarization drift that occurred during the run. This was converted
to a wavelength error using an empirical calibration, and
the polarization error was added in quadrature to the regression error
calculated as in Fig.~2.

\begin{figure}
\includegraphics[width=3.5in]{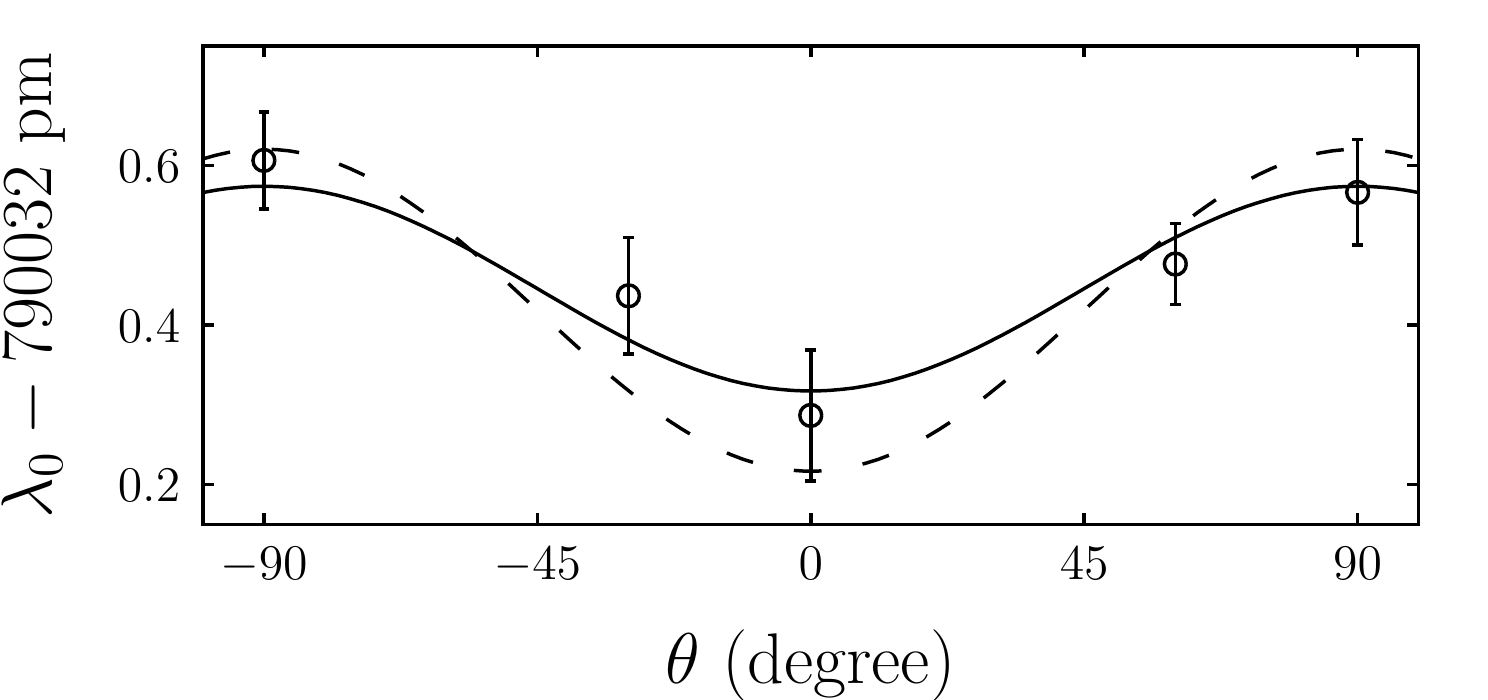}
\caption{
Effect of tensor polarizability. Data points show tune-out wavelengths
$\lambda_0$, as a function of the angle $\theta$ between the
linear polarization of the Stark beam and the $x$ axis of the trap. 
Each point is an average of several measurements performed as in 
Fig.~2. For each measurement $i$ the linear fit error is combined with
the estimated polarization error described in the text. The error
bars shown are then calculated as 
$\sigma^2 = 1/\sum_i \sigma_i^{-2}$.
The solid curve is a sinusoidal fit with a variable
offset and amplitude. The dashed curve is a fit with the amplitude
constrained to the expected value.}
\end{figure}

The tensor term in \eqref{alpha_terms} gives rise to a dependence
on the angle of the linear light polarization with respect to the
trap field, which can be seen in Fig.~3.
The polarization was adjusted using the
half-wave plate in the Stark beam.
For our geometry, the polarization angle
$\theta$ is related to $\xi$ in \eqref{alpha_terms}
via $\langle \cos^2\xi\rangle = 0.5 \cos^2\theta$,
where the brackets denote a time average for the magnetic field.

Near the tune-out wavelength, $\alpha$ and $\alpha^{(0)}$ can be
accurately approximated as linear functions
$(d\alpha/d\lambda) (\lambda-\lambda_0)$ and
$(d\alpha^{(0)}/d\lambda)(\lambda-\lambda^{(0)})$ respectively.
Here $\lambda_0$ is the measured value
shown in Fig. 3 and $\lambda^{(0)}$ is the desired zero of the scalar
term. The tensor contribution to $d\alpha/d\lambda$ is negligible,
so the two derivatives are nearly equal.
If we use this and set \eqref{alpha_terms} to zero, we obtain
\be
\lambda_0(\theta) = \lambda^{(0)} - \frac{\alpha^{(2)}}{d\alpha^{(0)}/d\lambda}
\left(\frac{3}{4}\cos^2\theta - \frac{1}{2}\right)
\ee
in the case of $\mathcal{V}\langle \cos\chi \rangle = 0$.
Fitting to this form, we obtain $\lambda^{(0)} = 790.032439(35)$ nm and
$\alpha^{(2)}/(d\alpha^{(0)}/d\lambda) = 390(120)$ fm. 
This fit is shown as the solid
curve in Fig. 3.

Alternatively, $\alpha^{(2)}$ and $d\alpha^{(0)}/d\lambda$ are almost
entirely due to contributions from the 5P manifold, and can be calculated
relatively precisely. The derivative term can be determined from 
\cite{Kien13}
\be
\begin{split}
\label{alpha5P}
\alpha^{(0)}_{5P} = 
 \frac{10}{\hbar\sqrt{15}} & \sum_{J',F'} 
\frac{ \left|d_{J'}\right|^2 \omega'}{\omega'^2-\omega^2}
\left(-1\right)^{1+F'} 
(2F'+1)\\
& \times \left\{ \begin{array}{ccc}
2 & 1 & F' \\
1 & 2 & 0
\end{array}\right\}
\left\{ \begin{array}{ccc}
F' & 3/2 & J' \\
1/2 & 1 & 2
\end{array}\right\}^2\\
\end{split}
\ee
for the $F=2$ ground state. 
Here the sum is over the angular momentum
quantum numbers of the 5P states, $\omega'$ is the transition
frequency to the $\ket{J',F'}$ state,
and $d_{J'} = \langle 5P_{J'} || d || 5S_{1/2} \rangle$ is the reduced
dipole matrix element. The $d_{J'}$ are known to about 500 ppm precision
from lifetime and photoassociation measurements
\cite{Volz96,Simsarian98,Gutterres02}.

Similarly, the tensor term is given by
\cite{Kien13}
\be
\begin{split}
\alpha^{(2)}_{5P}  =  \frac{20}{\hbar\sqrt{21}}
& \sum_{J', F'} 
\frac{|d_{J'}|^2 \omega'}{\omega'^2-\omega^2}
(-1)^{F'} (2F'+1) 
\\
& \times \left\{ \begin{array}{ccc}
2 & 1 & F' \\
1 & 2 & 2
\end{array}\right\}
\left\{ \begin{array}{ccc}
F' & 3/2 & J' \\
1/2 & 1 & 2
\end{array}\right\}^2.
\end{split}
\ee
Evaluating the ratio gives $\alpha^{(2)}/(d\alpha/d\lambda) = 538.5(4)$~fm, 
which is larger
than the value determined from our fit by about $1.3\sigma$.
If we constrain the fit to use the 
calculated value for $\alpha^{(2)}/(d\alpha/d\lambda)$,
we obtain $\lambda^{(0)} = 790.032388(29)$ nm,
about $1\sigma$ different from
the unconstrained result.
The constrained fit gives a $\chi^2$/d.o.f.\ of 1.2, compared to
0.5 for the unconstrained fit, both of which are reasonable. 
Since the calculated value for
$\alpha^{(2)}/(d\alpha/d\lambda)$ is expected to be accurate, we report 
the value obtained from the constrained fit.

\section{Error Estimation}

As noted, each run of the experiment yields a statistical error derived from
the linear fits of $\phi$ vs.\ intensity, and a polarization error based
on the measured polarization drift between the start and end of the run.
Each run takes several hours, so we are not confident that the polarization
change is linear, or even monotonic, throughout the run. We therefore use 
the full value of the polarization drift as an error estimate. The
polarization drift is in fact the largest error source in the 
measurement. The average polarization drift error is 126 fm, compared to 
the average statistical error of 60 fm. Averaging over the 21 measurements
used would reduce these values by $\sqrt{20}$.
However, both errors vary considerably
from run to run, so for the analysis we combine
the two errors for each data point in Fig. 3. The resulting fits have
the uncertainties cited above.

Another error contribution is the calibration
uncertainty in our wavelength measurement. 
We used a Bristol Instruments model 621A wave meter that displayed digits to 
1 fm, with results repeatable to about 10 fm.
We tested the meter by measuring four known saturated absorption lines
in K, Rb, and Cs. The results 
indicated a calibration correction of -40(5) fm at a wavelength of 790 nm.
This correction was applied to the data reported here.
The full wavelength calibration was performed both at the start and end of
data collection, and the two Rb lines were checked periodically throughout
the experiment.  No significant differences were observed.

A significant source of error is asymmetry in the Stark laser spectrum
\cite{Holmgren12,Henson15}.
The laser diode source produces broadband ASE light \cite{Chow90}. This
could be observed through its effect on the spontaneous emission
rate of the atoms, and indicated a background spectral density near
an atomic resonance of 
$S \approx P\times 10^{-17}$~Hz$^{-1}$, in terms of the total Stark power $P$. 
This is large enough to shift $\lambda_0$, depending
on the spectral distribution.
We controlled the effect by
by spectrally filtering the beam using a diffraction grating and pinhole.
Using a 0.4 nm $\approx 200$ GHz filter bandwidth, and assuming
a 10\% variation of the spectrum across that bandwidth, the estimated
spectral density would produce a shift of about 0.1 fm. However,
it is possible that the spectral density near 790 nm is 
larger than that at the atomic resonance. Such low spectral power levels
are difficult to measure directly, so we quantified the 
effect by comparing our $\lambda_0$ results obtained with 0.2 nm
and 0.4 nm filter bandwidths. 
About half our data was taken in each configuration. 
No measurable difference was observed, within our 30 fm precision. 
The expected error would scale as the bandwidth squared, indicating that the 
error for the smaller bandwidth configuration was less than 10 fm.
We use this as the uncertainty from the effect, 
though we expect it is an overestimate.
Asymmetry in the tails of the 
laser line itself could similarly shift the measurement, but this could be
ruled out at the 1-fm level using an optical spectrum analyzer.

Uncertainty in the trap magnetic field can affect our result by
changing the value of $\langle \cos^2\xi \rangle$ in the tensor term. The 
most significant effect is if the magnitude of the bias field varies
as it rotates. We were able to place a limit of 2\% on such variations
by measuring the Zeeman linewidth of the trapped atoms using rf spectroscopy.
In the worst case, this would induce a 5 fm shift on the 
value of $\lambda^{(0)}$. Other effects are smaller, 
including distortions from a dc background field of less than 1 G,
and angular misalignment of less than 3 degrees
between the Stark beam polarization 
measurement and the plane of the bias field.

The hyperpolarizability of the atoms characterizes the nonlinear Stark effect.
We estimate the effect by treating the $P_{1/2}$ and $P_{3/2}$ transitions
as two-level systems in the rotating wave approximation, and summing the
resulting energy shifts. At the tune-out wavelength, we 
obtain a net shift 
\be
\delta U \approx -\frac{|d_{1/2}|^4 \mathcal{E}^4}{32 \hbar^3 \Delta^3}
\ee
where $\Delta$ is the detuning from the $P_{1/2}$ transition and 
$\mathcal{E}$ is
the Stark field amplitude. At the maximum intensity used, this changes
$\lambda^{(0)}$ by only about 1 fm.

The effect of interatomic interactions is negligible, as the chemical
potential of the condensate is only about $2\pi\hbar\times 10$~Hz. The 
Zeeman shift from the trap field, however, is not small. By summing
the contributions of the individual Zeeman transitions, we calculate that it 
shifts the measured tune-out wavelength blue by 36 fm, so we have
added this amount to our reported values to give the estimated zero-field
result. From rf spectroscopy we know the bias field magnitude of 20.0(2) G
very accurately, so we estimate the error in this shift to be less than
1 fm.

Our error analysis results are summarized in Table \ref{errors}.
We sum the errors in quadrature to give our final reported 
one-sigma uncertainty of 32 fm.

\begin{table}
\caption{Estimated error contributions to $\lambda^{(0)}$. 
The entries for statistical error
and polarization drift report the average errors for each type, divided
by the square root of the number of measurements. In the analysis,
both errors were combined at each data point to give the reported 
combined error in the result. \label{errors}}
\begin{ruledtabular}
\begin{tabular}{lr}
\hspace{1em} Source & Error (fm) \\ \hline
Statistical & 13  \hspace{1em} \\
Polarization drift & 28 \hspace{1em} \\
\hspace{1em} Stat and polz. combined \hspace{2em} & 29 \hspace{1em} \\
Broadband spectrum & 10 \hspace{1em} \\
Wavemeter calibration & 5 \hspace{1em} \\
Trap field variation & 5 \hspace{1em} \\
dc background field & 2 \hspace{1em} \\
Hyperpolarizability & 1 \hspace{1em} \\ \hline
\hspace{1em} Total & 32 \hspace{1em}
\end{tabular}
\end{ruledtabular}
\end{table}

\section{Comparison to Theory}

One other experimental measurement of this tune-out wavelength exists, 
by Lamporesi {\em et al.}\ who obtained 790.018(2) nm \cite{Lamporesi10}. 
Our result is in considerable ($7\sigma$) 
disagreement, but those authors did not
report any special effort to control the light polarization. 
We expect therefore that their result 
is for the particular combination of scalar and vector polarizabilities
that was relevant to their experiment.

We can however make a useful comparison to theory. We first describe how
the theoretical result was obtained.
In the decomposition of Eq.~\ref{alphaparts}, the core terms
$\alpha_c$ and $\alpha_{vc}$ are 
are approximately static and are calculated
in the random-phase approximation \cite{Safronova11}.
The valence term for the 5S state can be expressed in atomic units as
 \begin{equation}
\alpha _{v}(\omega
)=\frac{1}{3}\sum_{k}\frac{{\left\langle
k\left\| d\right\| 5S\right\rangle }^{2}(E_{k}-E_{5S})}{(E_{k}-E_{5S})^{2}-%
\omega ^{2}},  \label{eq-pol}
\end{equation}
where $k=nP_{1/2}$ and $nP_{3/2}$. Up to $n=12$ we evaluate discrete
terms in this sum using experimental values for the state energies $E$. 
Experimental matrix elements from Ref.~\cite{Volz96} 
are used for the $5S-6P$ transitions while 
all other matrix elements use the all-order calculations of 
\cite{Safronova11}. 
The details of the methods are discussed in \cite{Safronova07}. 
While experimental values are available for the $5S-5P$ matrix elements
\cite{Volz96},
the theoretical values are estimated to have a more accurate ratio,
which is most important here.
For $n>12$, the remaining `tail' contributions are calculated in the 
Dirac-Hartree-Fock approximation.
The state energies and matrix elements are listed in Table \ref{tab1}. 
Using these values, the tune-out wavelength is predicted to lie at
$\lambda^{(0)} = 790.02568$~nm as indicated. 

\begin{table}
\caption{Breakdown of the contributions to the $5S$ polarizability in 
Rb at $\lambda=790.02568$~nm.
Reduced matrix elements $d$ and polarizability contributions are given in 
atomic units. Experimental matrix elements from Ref.~\protect\cite{Herold12} 
are used for the  $5S-6P$ transitions; remaining matrix elements are from the 
all-order calculations \protect\cite{Safronova11,Herold12}. Uncertainties 
are given in parenthesis.  Experimental energies $\Delta E$
are measured from the ground state and given in cm$^{-1}$
\protect\cite{Kramida13}.} 
\begin{ruledtabular}
\begin{tabular}{lrrr}
\multicolumn{1}{c}{Contr.}& \multicolumn{1}{c}{$\Delta E$}&
\multicolumn{1}{c}{$d$}& \multicolumn{1}{c}{$\alpha_0$} \\ \hline
$5P_{1/2}	$	&	12578.951	&	4.2199		&-8233.6\\
$6P_{1/2}	$	&	23715.081	&	0.3235(9)	&0.451(3)	\\
$7P_{1/2}	$	&	27835.05	&	0.115(3)	&0.044(2)	\\
$8P_{1/2}	$	&	29834.96	&	0.060(2)	&0.011(1)		\\
$9P_{1/2}	$	&	30958.91	&	0.037(3)	&0.004(1)		\\
$10P_{1/2}	$	&	31653.85	&	0.026(2)	&0.002		\\
$11P_{1/2}	$	&	32113.55	&	0.020(1)	&0.001		\\
$12P_{1/2}	$	&	32433.50	&	0.016(1)	&0.001		\\
$(n>12)P_{1/2}	$	&		&			&	0.022(22)	\\ [0.5pc]
$5P_{3/2}	$	&	12816.54939	&	5.9550 	&	8222.9	\\
$6P_{3/2}	$	&	23792.591	&	0.5230(8)   &1.173(4)\\
$7P_{3/2}	$	&	27870.14	&	0.202(4)	&0.135(6)\\
$8P_{3/2}	$	&	29853.82	&	0.111(3)	&0.037(2)\\
$9P_{3/2}	$	&	30970.19	&	0.073(5)	&0.015(2)\\
$10P_{3/2}	$	&	31661.16	&	0.053(4)	&0.008(1)\\
$11P_{3/2}	$	&	32118.52	&	0.040(3)	&0.004(1)\\
$12P_{3/2}	$	&	32437.04	&	0.033(2)	&0.003	\\
$(n>12)P_{3/2} 	$	&		&			&	0.075(75)	\\
Core + vc		&		&			&	8.709(93)	\\
Total		    &		&			&	0.001		
\end{tabular}
\end{ruledtabular}
\label{tab1}
\end{table}

The uncertainty in the theoretical value is dominated by uncertainty in 
the 5P matrix elements.
In Table~\ref{tab2} we compare the matrix elements 
obtained using various approximations \cite{Safronova11}. All of the
methods are intrinsically relativistic.
The calculations in Table~\ref{tab2} of 
the tune-out wavelength differ only in the
values of these two matrix elements, with all other values taken from 
Table~\ref{tab1}.

\begin{table*}
\caption{Reduced electric-dipole matrix elements for 
the $5S-5P_{J}$ transitions \cite{Safronova11}, 
values of the tune-out wavelength $\lambda^{(0)}$, 
the matrix element  ratio $R$. 
Theoretical methods:
DF is the lowest-order Dirac-Hartree-Fock, II and III are 
second- and third-order many-body perturbation theory values,  
SD and SDpT are \textit{ab initio} all-order values calculated in the
single-double approximation and with inclusion of the partial 
triple contributions, and SD$_{sc}$, SDpT$_{sc}$ are 
corresponding scaled all-order values. 
Experimental values are averages of several experimental 
measurements \protect\cite{Volz96,Simsarian98,Gutterres02}.}
\begin{ruledtabular}
\begin{tabular}{lrrrrrrrr}
\multicolumn{1}{c}{}& \multicolumn{1}{c}{DF}&
\multicolumn{1}{c}{II}& \multicolumn{1}{c}{III}&
\multicolumn{1}{c}{SD}& \multicolumn{1}{c}{SD$_{sc}$}&
\multicolumn{1}{c}{SDpT}& \multicolumn{1}{c}{SDpT$_{sc}$}&
\multicolumn{1}{c}{Expt.}\\
 \hline
$5S-5P_{1/2}$	& 4.8189	&    4.5981	&   4.1855&	   4.2199&	4.2535	&   4.2652	 &   4.2498	 &   4.233(2)  \\
$5S-5P_{3/2}$	& 6.8017	&    6.4952	&   5.9047&	   5.955&	6.0031	&   6.0196	  &  5.9976	  &  5.978(4)\\
$\lambda^{(0)}$	(nm) & 790.02603	&790.03155  &790.02380&  790.02568&	790.02636&  790.02632&	790.02607&	790.031(6)\\
$R$	        & 1.9922	&    1.9954	&   1.9902&	   1.9914&	1.9919	 &  1.9918	  &  1.9917	  &  1.995(3)\\
\end{tabular}
\end{ruledtabular}
\label{tab2}
\end{table*}

\begin{table}
\caption{Reduced electric-dipole matrix elements for the $5S-5P_{J}$ transitions and the corresponding line strength
  ratio $R$.}
\begin{ruledtabular}

\begin{tabular}{lrrr}
\multicolumn{1}{c}{}& \multicolumn{1}{c}{DF}&
\multicolumn{1}{c}{DF+Breit}& \multicolumn{1}{c}{DF+QED}\\
 \hline
 $5S-5P_{1/2}$	&	4.8189&	4.8192&	4.82038 \\
 $5S-5P_{3/2}$	&	6.8017&	6.8023&	6.80384 \\
 $R$	        &   1.9922&	1.9923&	1.9923 \\
\end{tabular}
\end{ruledtabular}
\label{tab3}
\end{table}

The most accurate methods are expected to be the 
four all-order calculations 
SD, SDpT, SD$_{sc}$, and SDpT$_{sc}$.
We take the average of these as the final theoretical values,
and use them to calculate
$\lambda^{(0)} =790.0261(7)\textrm{nm}.$ 
The uncertainty is estimated from the spread in the four values.
While the scaling (SD$_{sc}$, and SDpT$_{sc}$) technique
is supposed to account for a class of missing correlation effects, 
the scaling affects only about half of the
correlation correction in this transition. 
Therefore we use the full spread of the values as our error estimate
to allow for the effects missed by scaling.
We note that this uncertainty estimation is 
approximate since we are attempting to account 
for unknown correlation effects due to triple, quadrupole,
and higher excitations.

The estimated uncertainty in $\alpha$ 
from all of the non-5P contributions is about 0.12 au. Via the derivative
$d\lambda/d\alpha = -397$ fm/au, this leads to a wavelength error of 50 fm,
about ten times smaller than the uncertainty from the 5P levels. 
The net value of the non-5P contributions does give a significant shift
of -4.2 pm, mainly from the core polarizability.

The wavelength value determined above does not include
the effects of hyperfine structure.
This can be incorporated using Eq.~\eqref{alpha5P} for the 5P levels,
using the theoretical estimate for the dipole matrix elements.
The effect of hyperfine structure from all other levels is negligible.
This yields $\lambda^{(0)} = 790.0312(7)$~nm,
in reasonable agreement with the experimental
value of 790.032388(32)~nm. The values differ by $1.7\sigma$, with
the theoretical uncertainty about twenty times larger than that of the 
experiment.

As noted,
the 5P matrix elements themselves contribute primarily through their ratio
\begin{equation}
R=\frac{|\langle 5P_{3/2}||d||5S\rangle|^2}{|\langle 5P_{1/2}||d||5S\rangle|^2}.
\label{eq1}
\end{equation}
This is useful, because the theoretical accuracy of the ratio is better
than that of the individual matrix elements since a large fraction of
the correlation corrections cancel. 
This can be seen in the calculations in Table~\ref{tab2}. Using
the same error estimation procedure as above, we obtain a ratio
$R=1.9917(5)$.

None of the matrix element values in Table~\ref{tab2} include 
Breit or QED corrections.
We evaluate the importance of these effects 
in the lowest-order DF approximation and 
summarize the resulting values in Table~\ref{tab3}.
First, we carry out the DF calculation with 
the Breit interaction included on the same 
footing with the Coulomb interaction 
(see, for example, Ref.~\cite{Dzuba06}). 
The resulting values are listed in the column labeled ``DF+Breit.'' 
Then, we carry out the DF calculation with the inclusion of the QED 
model potential, constructed as described in Ref.~\cite{Flambaum05}. 
The Breit interaction is excluded in this calculation to separate the 
two effects. 
We find that both Breit and QED corrections are five times smaller than our 
uncertainty in the correlation contribution to the ratio. However,
we include the shifts in our estimate $R = 1.9919(5)$.

An experimental determination of the matrix element ratio requires some
theoretical input  \cite{Holmgren12}. 
The scalar polarizability can be expressed
\be
\alpha^{(0)} = A + |d_{1/2}|^2\left(K_{1/2} + K_{3/2} R \right)
\ee
where $A$ includes $\alpha\sub{c}$, $\alpha\sub{cv}$, and contributions 
from valence states above $5P$. Using the values from Table \ref{tab1} gives 
$A = 10.70(12)$ au.
The experimental value for 
$d_{1/2}$ is 4.233(2) au \cite{Volz96,Simsarian98,Gutterres02}.
The coefficients $K_{J'} \equiv \alpha^{(0)}_{5P_{J'}}/|d_{J'}|^2$ 
can be obtained from Eq.~\eqref{alpha5P} and
our result for $\lambda^{(0)}$.
Setting $\alpha^{(0)} = 0$ and solving for
$R$ yields $1.99221(3)$.
This differs from our theory result by $0.6\sigma$, and is about twenty
times more accurate. Both values are consistent with 
the ratio of the previous 
experimental matrix elements, $R = 1.995(3)$.

\section{Conclusions}

\begin{figure}
\includegraphics[width=3.5in]{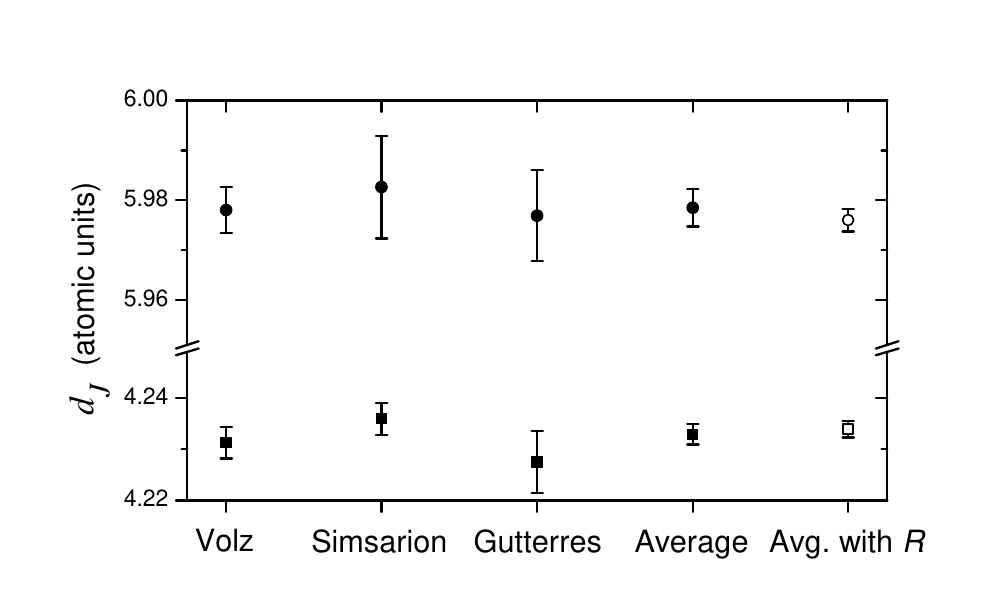}
\caption{
Matrix element values. The points show various determinations of
the $d_J = \langle 5P_J || d || 5S_{1/2} \rangle$ matrix elements,
for $J = 1/2$ (squares) and $J=3/2$ (circles).
The first three points are measurements by Volz {\em et al.}\ 
\protect\cite{Volz96},
Simsarian {\em et al.}\ \protect\cite{Simsarian98} and Gutterres {\em et al.}\
\protect\cite{Gutterres02}. The
fourth point is the error-weighted average from the three groups.
The fifth point (hollow) is the error-weighted average of all six
measurements, with the constraint $d_{3/2}^2/d_{1/2}^2 = R = 1.99219$
obtained in the present work.
\label{exp_elements}
}
\end{figure}

Our measurement of the $^{87}$Rb 790 nm tune-out wavelength illustrates
that tune-out wavelength spectroscopy can provide high precision information
about atomic matrix elements. As one immediate application, our 
measurement of the matrix element ratio provides a moderate improvement to the
absolute values of the $5P_{1/2}$ and $5P_{3/2}$ matrix elements.
Each of these elements has been determined with about 0.1\% precision
in three previous investigations \cite{Volz96,Simsarian98,Gutterres02}.
Our 15 ppm determination of the ratio allows all six measurements
to be combined, reducing the total estimated error in each
element by about a factor of $\sqrt{2}$, as seen in
Fig. \ref{exp_elements}. The resulting best values are
$d_{1/2} = 4.2339(16)$ and $d_{3/2}=5.9760(23)$.
Precise knowledge of $R$ may permit yet further improvements using the technique
of Ref.~\cite{Derevianko02}.

Our results have several important conclusions in regards to the 
atomic theory calculations.
First, the good agreement between the measured and calculated values of
$\lambda^{(0)}$ provides confirmation of the theoretical accuracy.
Prior to our measurement, the theory result was about five times more accurate
than the best experimental estimate, making the theoretical prediction
difficult to check. In particular, our result validates the procedure
used to estimate the theoretical error, since the error accurately reflects
the disagreement with experiment. This type of error validation is valuable
since theoretical error estimates are both challenging and important to obtain.

Second, we demonstrate 
that the ratio of matrix elements can be a useful measure of the 
accuracy of theoretical approaches to include electron correlations.
Note that the second-order and third-order values in Table~\ref{tab2}
are outside of the theory uncertainty estimate, 
and disagree significantly with the experimental
result. These methods are thus confirmed to be less accurate than the all-order
techniques. 
    
Third, the accuracy of the experimental ratio value is sufficient to test 
the Breit and QED effects if a more accurate treatment of correlations
is carried out. It may be possible to achieve this in the
full triple coupled-cluster approach used 
to treat Cs parity violation \cite{Porsev09}.
If successful, this would help support the theoretical methods and
thus clarify the parity violation results  \cite{Flambaum05,Dzuba06}.

The method we have demonstrated can readily be applied to other
tune-out wavelengths in Rb, which we hope to pursue in future work. 
We hope in this way that the Rb atom can be established as a well-known
reference atom for testing theoretical techniques. The significant
advance in experimental precision should provide a useful benchmark
for some time to come.

More generally our method can be applied to any Bose-condensed atomic species,
which includes many species used in precision measurement applications.
We hope that the improved knowledge of matrix elements made possible will
prove valuable.

\begin{acknowledgments}
We are grateful to A. Cronin for helpful discussions and comments on 
the manuscript, and to V. Dzuba for the use of his QED code.
The Virginia group was supported by the National Science Foundation Grant
No.\ 1312220, NASA Grant No.\ 1502012, and the Jefferson Scholars Foundation. 
This research was performed in part under the sponsorship of the US Department
of Commerce, National Institute of Standards and Technology.
\end{acknowledgments}


\end{document}